\documentstyle[psfig]{elsart}

\newcommand{\bd}[1]{ \mbox{\boldmath $#1$}  }
\newcommand{\gtrsim} {  \stackrel{\mbox{\large $_>$}}
                                 {\mbox{\large $_\sim$}}  }
\newcommand{\lesssim} { \stackrel{\mbox{\large $_<$}}
                                 {\mbox{\large $_\sim$}}  }

\begin{document}
\begin{frontmatter}

\title{Inclusive quasielastic electron scattering on $^6$He: 
a probe of the halo structure}

\author{E. Garrido and E. Moya de Guerra}
\address{Instituto de Estructura de la Materia, CSIC, Serrano 123,
 E-28006 Madrid, Spain}
\date{\today}

\maketitle

\begin{abstract} 

We investigate inclusive electron scattering reactions on two-neutron 
halo nuclei in the quasielastic region. Expressions for the
cross section and structure functions are given assuming that the halo 
nucleus can be described as a three-body system ($\mbox{core}+n+n$).
The method is applied to $^6$He. We compute cross sections and
structure functions, and investigate the kinematic conditions 
for which the observables are determined either by $\alpha$-knockout
or by halo neutron-knockout. The
optimal kinematical domain to disantangle the momentum
distributions of the various components of the three--body system
($q \lesssim 200$ MeV/c and $\omega < q^2/2M_N + 20$ MeV) are explored.
\vspace{1pc}
\end{abstract}
\end{frontmatter}

\par\leavevmode\hbox {\it PACS:\ } 25.30.Fj, 25.10.+s, 21.45.+v

\section{Introduction}

Halo nuclei are the most intriguing objects among recent nuclear
physics discoveries. Their small binding energies and large spatial
extensions had no precedent and were totally unexpected in the
realm of  systems governed by the nuclear force. Borromean two--neutron
halo systems show the additional peculiarity that the three--body
(core+$n$+$n$) system completely falls apart when either of the
three constituents drops out. Among the latter $^6$He ($\alpha$+$n$+$n$)
is the one that is most stable and has been extensively studied.
Since the initial discovery by Tanihata and collaborators \cite{tan85,tan88},
much progress has been made in  theoretical \cite{han87,joh90,han95,zhu93} 
and experimental studies of halo nuclei. The available experimental information 
comes basically from nucleus--nucleus collisions
\cite{ann90,orr92,orr95,zin95,nil95,hum95,zin97}. 
This experimental information has the inherent difficulty of 
disentangling nuclear reaction mechanisms
from nuclear structure effects. This is why the idea of using nonhadronic 
probes to obtain complementary information on halo nuclei has been 
put forward.

The aim of this paper is to point out that, as for the case of
stable nuclei, detailed information on charge density and
momentum distributions in these systems can be obtained with electron
beams. To this end we show theoretical results on  
differential cross sections and structure functions for the case of
electron scattering from $^6$He.  We do not discuss here important
technical difficulties stemming from the short lives of halo
nuclei and small production rates, which work against the feasibility
of experiments involving measurements of cross sections for
electron scattering from a halo--nucleus secondary beam. 
We just mention that in spite of
these difficulties such experiments are in project at RIKEN
\cite{ohk97}, and are also being considered at GSI and
Mainz.

To our knowledge, the first theoretical study of electron scattering
from halo nuclei was presented in ref.\cite{gar99}. In this reference
we developed the formalism focusing on elastic scattering and 
on exclusive $(e,e^\prime \alpha)$ and $(e,e^\prime n)$ coincidence
experiments. In particular we studied the reactions 
$^6$He($e, e^\prime \alpha$)$nn$ and $^6$He($e, e^\prime n$)$\alpha n$
using a $^6$He three--body wave function that describes accurately
all presently known phenomenology. The results that we
present and discuss here are based on that previous work, where
details of calculations can be found. 

Determination of the spectral
function requires coincidence measurements in exclusive experiments.
Since the complexity of these experiments is high, it is natural to
think that inclusive $(e,e^\prime)$ experiments, in which only the
scattered electron is detected, will be the ones to be performed 
first. Moreover, it is difficult to select adequate kinematical
conditions for exclusive experiments due to the large number of
variables (energy and momentum transfer ($\omega$,$q$) and energy and
direction of the knocked out particle). In this regard it is 
mandatory to know first what are the ($\omega$,$q$) domains where
the quasielastic peaks for $\alpha$ knockout and halo neutron 
knockout dominate the total ($e$,$e^\prime$) differential cross 
section for the halo nucleus.  For the first time we show
here results on inclusive scattering over an ($\omega, q$) region
that allows to visualize the appearance of different quasielastic
$e$--$\alpha$ and $e$--$n$ peaks in different kinematical regions.  
We also show that coincidence measurements
in this separate kinematical regions allow to determine the spectral
functions of the core and the halo neutrons, and thus the detailed 
structure of the three--body wave function in momentum space.

\section{Model and method}

\begin{figure}[ht]
\centerline{\psfig{figure=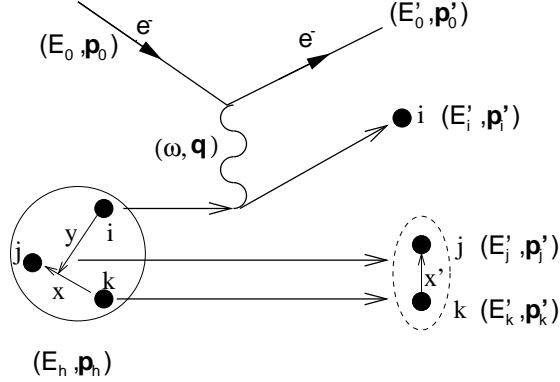,width=9.5cm,%
bbllx=-2cm,bblly=0.0cm,bburx=15.5cm,bbury=10.cm,angle=0}}
\caption[]{ Scheme and definition of the coordinates used.}
\label{fig1}
\end{figure}

To probe the structure of the three--body wave function of
two--neutron halo nuclei we consider the
reaction sketched in fig.\ref{fig1}.
The collision between the electron and the 
target is taken as a process in which one of the three constituents
in the target absorbs the whole energy and momentum transfer 
($\omega$,$\bd{q}$) and is knocked  out from the target.
Core excitations and core breakup reactions are not considered here, 
and the inclusive cross section 
\begin{equation}
\frac{d^3\sigma}{dE^\prime_0 d\Omega_{p^\prime_0}}=
\sigma_M \left(
V_L W_L + V_T W_T
\right)
\label{eq2}
\end{equation}
can then be computed as the sum of three
terms corresponding to the absorption of the virtual photon by each of the
three constituents in the halo nucleus. In the same way the structure functions 
$W_L$ and $W_T$ are given by the sum of the  elementary structure 
functions $W_{L,T}^{(i)}$ corresponding to a process where the constituent $i$
absorbs the virtual photon. In the particular case of $^6$He, since the
$\alpha$--core has zero spin, one has 
$W_L=2 W_L^{(n)} + W_L^{(\alpha)}$ and $W_T=2W_T^{(n)}$.
In the reaction shown in fig.\ref{fig1}
the incident and scattered electrons
as well as the ejected particle (constituent $i$) are described as plane waves 
(Plane Wave Impulse Approximation).
Since we are dealing with a borromean target the residual two--body system
made by particles $j$ and $k$ is in the continuum, and the
final interaction between them is included in the calculation.

Working in the frame of the halo 
nucleus ($\bd{p}_h=0$) the differential cross section of the process
shown in fig.\ref{fig1} is:

\begin{equation}
\frac{d^6\sigma^{(i)}}
   {dE^\prime_0 d\Omega_{p^\prime_0} d\Omega_{p^\prime_i} dE^\prime_x} =
(2 \pi)^3  \frac{p^\prime_i E^\prime_i (m_j+m_k)}{M_h} f_{rec}
\sigma^{ei}(\bd{q}, \bd{p}_i)  S(E^\prime_x,\bd{p}_i)
\label{eq1}
\end{equation}
where the different momenta and energies are shown in the figure.
$\bd{p}^\prime_x$ is the relative momentum between particles $j$ and
$k$ in the final state.
$m_j$, $m_k$ and $M_h$ are the masses of the halo constituents $j$,
$k$ and the total mass of the halo nucleus, respectively. 
$E^\prime_x=p^{\prime 2}_x/2\mu_{jk}$ is the kinetic energy of the
system made by particles $j$ and $k$ referred to its own center
of mass and $\mu_{jk}$ its reduced mass. 
$f_{rec}$
is the recoil factor, $\sigma^{ei}(\bd{q}, \bd{p}_i)$ is the cross
section for elastic electron scattering on constituent $i$, and
$S(E^\prime_x,\bd{p}_i)$ is the spectral function. 

The cross section $\sigma^{ei}(\bd{q}, \bd{p}_i)$ takes the general
form $\sigma^{ei}(\bd{q},\bd{p}_i)=\sigma_M (
   V_L {\cal R}^{(i)}_L+V_T{\cal R}^{(i)}_T+
   V_{LT}{\cal R}^{(i)}_{LT}+V_{TT}{\cal R}^{(i)}_{TT})$,
where the $V$'s are kinematic
factors and the ${\cal R}^{(i)}$$^\prime$s are structure functions
associated to an electron scattering process on constituent
$i$. When this constituent has spin 0 (as  
the $\alpha$ particle in $^6$He) only the longitudinal structure
function appears (see \cite{gar99} for details). 

$S(E^\prime_x,\bd{p}_i)$ is the spectral function, that
is interpreted as the probability for the electron to remove a particle
from the nucleus with internal momentum $\bd{p}_i$ leaving the
residual system with internal energy $E^\prime_x$.
This function contains the information about the nuclear structure,
in particular it contains the whole dependence on the halo nucleus wave
function and on the continuum wave function of the residual two--body
system. It is defined as the square of the overlap between both wave functions,
which reduces to the square of the Fourier transform of the halo 
wave function when the final interaction between the two
surviving constituents is neglected. 

The cross section for an inclusive quasielastic electron scattering 
process is obtained from eq.(\ref{eq1}) after integration over the 
unobserved quantities, i.e. $\Omega_{p^\prime_i}$ and $E^\prime_x$,
and after summation over the index $i$, that runs over all the three
constituents of the halo nucleus, the two neutrons and the core.
Since the $LT$ and $TT$ terms in the cross section automatically
disappear after integration over $\Omega_{p^\prime_n}$ \cite{cab93},
one immediately recovers the expression in eq.(\ref{eq2}).

In what follows we discuss results for $^6$He, that
is described as a system made by an $\alpha$-particle and
two neutrons. Compared to other halo nuclei $^6$He has the advantage 
that the interactions between the constituents are well known, and 
that the $\alpha$--core is tightly bound. Since the spin of the
core is zero we get for $^6$He
\begin{equation}
W_L=2W_L^{(n)}+W_L^{(\alpha)} ; \hspace*{1cm} W_T=2W_T^{(n)}
\end{equation} 
where
\begin{equation}
W^{(i)}_{\kappa}= (2 \pi)^4 \frac{m_j+m_k}{M_h} 
\int d\theta_{p^\prime_i} dE^\prime_x 
p^\prime_i E^\prime_i f_{rec} \sin \theta_{p^\prime_i}
{\cal R}^{(i)}_{\kappa}(\bd{q}, \bd{p}_i) S(E^\prime_x, \bd{p}_i)
\label{eq3}
\end{equation}

The neutron structure functions ${\cal R}_L^{(n)}$ and ${\cal R}_T^{(n)}$
are computed using the CC1
prescription for the nucleon current \cite{for83}, and
their analytic expressions are given in \cite{gar99}.
The structure function ${\cal R}_L^{(\alpha)}$ is the
square of the Fourier transform of the $\alpha$--particle
charge density. We use the experimental $\alpha$ charge density 
parameterized as a sum of two gaussians as in table V of \cite{vri87}. 
Explicit expressions
of the spectral functions $S(E^\prime, \bd{p}_n)$ and 
$S(E^\prime, \bd{p}_\alpha)$ are given by eqs.(33) and (56) of
\cite{gar99}.
We use a three--body $^6$He wave function obtained by solving the Faddeev
equations in coordinate space. The procedure is shown in
detail in \cite{fed95}. The continuum wave function of the residual
two--body system in the final state is written as a partial wave 
expansion \cite{gar97} whose radial functions are obtained by solving the
Schr\"{o}dinger equation with the corresponding neutron--neutron or 
neutron--$\alpha$ potential.
For the $n$-$n$ interaction
we use a simple potential with gaussian shape including a central,
spin--orbit, spin--spin, and tensor terms, whose parameters are
adjusted to reproduce low energy neutron--neutron scattering data.
For the $n$-$\alpha$ interaction we use a potential with central
and spin--orbit terms also with gaussian shapes and parameters adjusted
to reproduce the phase shifts for $s$, $p$, and $d$ waves up to
20 MeV. These potentials provide an accurate description of the
available information on $^6$He. Details about these potentials are given 
in \cite{gar97b}.

\section{Results}

\begin{figure}
\centerline{\psfig{figure=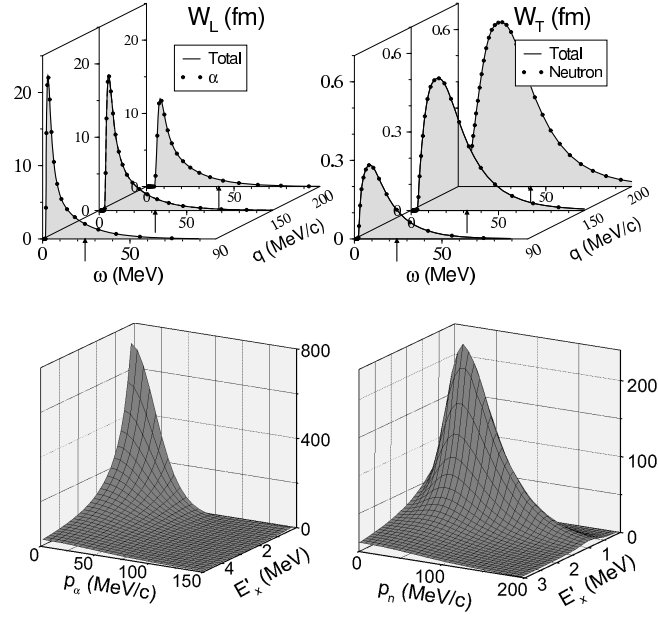,width=6.0cm,%
bbllx=6.5cm,bblly=5.cm,bburx=17.0cm,bbury=25.cm,angle=0}}
\vspace{-2.3cm}
\caption{ Upper part: $W_L$ and $W_T$ for momentum transfers
$q$=90, 150 and 200 MeV/c as a function of $\omega$.
The dotted lines show the contribution
from the $\alpha$--particle in $W_L$ and from the halo neutrons
in $W_T$. The arrows indicate the $\omega$-values where core
breakup contributions may show up. Lower part: Spectral functions
for $\alpha$--knockout (left) and halo neutron knockout (right).}
\label{fig2}
\end{figure}

In the upper part of fig.\ref{fig2} we show the longitudinal
and transverse structure functions $W_L$ and $W_T$ for varying 
energy transfer $\omega$ at fixed values of the momentum 
transfer ($q$=90, 150 and 200 MeV/c). Note that for
each $q$ the accessible $\omega$ values are constrained to 
be less than $q$ ($\omega^2 - q^2 <$0).
In the upper--left part of fig.\ref{fig2} we see that $W_L$ is
dominated by the contribution of the $\alpha$--knockout.
The latter is seen as a strong peak centered at $\omega \approx
q^2/2m_\alpha$, to which we refer as the $\alpha$--peak. The
contribution from the halo neutrons to $W_L$ is not visible at
the ($\omega$,$q$) values considered in this figure. For
$q>200$ MeV/c the $\alpha$--peak decreases while the contribution
from the halo neutrons increases with increasing $q$ as the $\alpha$ and 
neutron form factors, respectively. For $q$--values beyond
250 MeV/c the $\alpha$--peak gets shorter and shorter and the 
contribution to $W_L$ from halo neutrons shows up at
$\omega \approx 30$ MeV. However, at this energy and
momentum transfer this contribution may have a large overlap with 
the contribution from core breakup, i.e., from one--nucleon
knockout from the $\alpha$--particle. We have indicated
with vertical arrows the $\omega$--values where core breakup
contributions may show up more. For $\omega$--values to
the left of the arrows $W_L$ is dominated by the $\alpha$--peak.

The structure function $W_T$ is shown in the upper--right
part of fig.\ref{fig2}. As already announced, only the halo 
neutrons contribute to these peaks (we refer to them as
halo--neutron peaks) which are seen to increase with 
increasing $q$--values and are centered at 
$\omega$$\approx$$q^2/2m_n$. Again we indicate by arrows the 
$\omega$--values where core breakup contributions to $W_T$ will
show up more. Core breakup contributions to both $W_L$ and
$W_T$ are expected to be more sizeable at $q$$>$250 MeV/c and
$\omega$$>$25 MeV. Typically data on one nucleon knockout
from a free $\alpha$--particle are available at larger
$q$--values than those considered in fig.\ref{fig2}
\cite{dyt88}. An estimate of the
core breakup contributions to the $W_L$ and $W_T$ structure
functions for inclusive $^6$He($e,e^\prime$) can be obtained
from this reference for 280 MeV/c $\lesssim q \lesssim $ 720
MeV/c. The maximum values of $W_L$ and $W_T$ for 
$^4$He($e,e^\prime$) extracted from this
reference are $W_L$$\sim$1.5 fm and $W_T$$\sim$2 fm, which
take place at $q$$\sim$360--380 MeV/c and 
$\omega$$\sim$70--80 MeV. 

Therefore the low ($\omega$,$q$) domain shown
in the upper plots of fig.\ref{fig2} appears to be
optimal to obtain independent and reliable measurements
of the $\alpha$ and the halo neutron peaks, and thus on
the momentum distributions of the different components
of the three--body system. Indeed, as seen in fig.\ref{fig2},
$W_L$ provides information about the spectral function
$S(E^\prime_x, p_\alpha)$, corresponding
to an $\alpha$--knockout process, while $W_T$ contains
information about the spectral function $S(E^\prime_x, p_n)$,
corresponding to a halo neutron knockout process. Both spectral
functions are shown in the lower part of the figure. 
From the plots one could naively think that $W_T$ is wider than $W_L$ 
because of the different widths of the corresponding spectral functions 
in the lower part of the figure. However the main reason for this is the 
different mass of core and neutron. The heavier the particle giving rise
to a structure function, the narrower the structure function. Since
the $\alpha$-particle is four times heavier than the neutron one
has that $W_L$ is narrower than $W_T$. More details of the spectral
functions are discussed in ref.\cite{gar99}. They could
be measured by coincidence experiments $^6$He $(e,e^\prime \alpha)2n$
and $^6$He$(e,e^\prime n)^5$He in the low ($\omega$,$q$) region
explored here.

\begin{figure}
\centerline{\psfig{figure=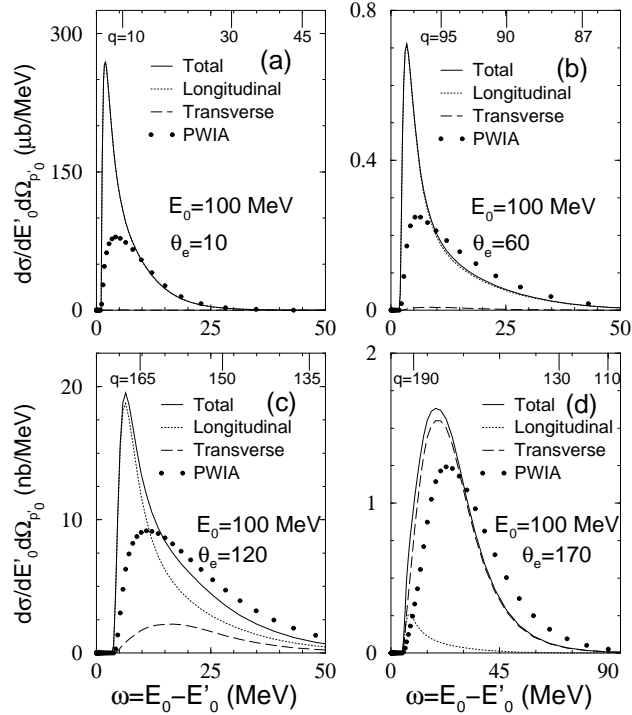,width=4.5cm,%
bbllx=5cm,bblly=4.cm,bburx=15.5cm,bbury=24.cm,angle=0}}
\vspace{1.5cm}
\caption[]{ Differential cross section for $E_0$=100
and various scattering angles.
The solid lines, short-dashed lines and long-dashed lines
are the total cross section, the longitudinal contribution
and the transverse contribution, respectively. The dotted
line gives is a full PWIA calculation.
The $q$--values in the upper axes are given in MeV/c.}
\label{fig3}
\end{figure}

To show that indeed we can get separately information of the
$\alpha$ and halo--neutron peaks, we show in fig.\ref{fig3}
how $W_L$ and $W_T$ enter in the inclusive differential
cross section for incoming energies ($E_0$) and scattering
angles ($\theta_e$) adequate to explore the low ($\omega$,$q$)
domain discussed above. In this figure we show the inclusive
$^6$He($e,e^\prime$) cross section for $E_0$=100 MeV, and
four different values of $\theta_e$. There is a clean 
separation of longitudinal and transverse contributions
as one goes from forward angles ($\theta_e$=10$^\circ$,
60$^\circ$) to backward angles ($\theta_e$=120$^\circ$,
170$^\circ$). The results are plotted as a function of
$\omega$. For fixed $E_0$ and $\theta_e$ values $\omega$
and $q$ are linked by the relation 
$q^2=\omega^2+4E_0(E_0-\omega)\sin^2(\theta_e/2)$. To guide
the reader $q$--values are indicated in the upper horizontal
axes. Similar plots are obtained if one varies the incoming
energy in the range 70 MeV $\lesssim E_0 \lesssim$ 150 MeV.
At forward angles only the $\alpha$--peak ($W_L$) contributes,
while as $\theta_e$ increases beyond 90$^\circ$ the halo
neutron contribution ($W_T$) starts to increase till it becomes
the dominant contribution to the differential cross section.
At 180$^\circ$ there is no contribution from the longitudinal
structure function and the only contribution remaining is 
the transverse one. This is a general property that is well
known and prevails for electron scattering from any target
(see for instance ref.\cite{moy86} and references therein).
Not included in these figures is the contribution from elastic 
electron--$^6$He scattering, that was considered in \cite{gar99}
and contributes only at forward angles and $\omega\rightarrow 0$.
We can see in fig.\ref{fig3} that, in each plot,  the peak of the 
differential cross section falls in an 
($\omega$,$q$) range that is expected to be free from
core breakup: $q\lesssim$ 200 MeV/c and $\omega < q^2/2M_N
+$$S_N$, where $M_N$ is the nucleon mass and $S_N$$\approx$20 MeV 
is the nucleon separation 
energy in the $\alpha$--particle. The latter process will be 
considered subsequently.
In the calculations only the final interaction between particles
$j$ and $k$ (see fig.~\ref{fig1}) is included. The important role 
that this interaction plays can be appreciated in fig.\ref{fig3}
by comparison to PWIA results (dotted line), where this interaction
is switched off. The difference between both calculations is due to
the structure that the two--body spectator resonances create in 
the spectral functions. It would be interesting to study the effect
caused by the final state interaction between the knocked out
particle and the spectators. The treatment of three--body interactions
in the final state is highly non--trivial and deserves further work.
In particular this requires knowledge of the three--body resonance
states. Experimental knowledge of three--body resonances is scant,
but progress along these lines is being made \cite{aum99}.

\section{Conclusion}

In summary, we have investigated inclusive electron scattering
on two-neutron halo nuclei, and particularly on $^6$He, to explore
kinematical regions where the quasielastic processes
$^6$He$(e,e^\prime n)\alpha n$ and $^6$He$(e,e^\prime \alpha)n n$
are favored. Considering the halo nucleus as a three--body (core+$n$+$n$)
system, the differential cross section is calculated as the sum of terms 
coming from halo--neutron knockout and core knockout. 
It is then desirable to determine under what conditions
the behavior of the different observables, cross section
and structure functions, is dominated
either by the halo neutrons or by the $\alpha$--core.

At forward scattering angles, for
low $q$ and $\omega$ ($q$$\lesssim$200 MeV/c, 
$\omega$$\sim$$q^2/m_\alpha$)
the behavior of the differential cross section and of the
longitudinal structure function is dictated  by the 
$\alpha$--knockout process. In this domain the cross sections
are large and dominated by the $\alpha$--peak.
It is therefore in this region where coincidence
$(e,e^\prime \alpha)$ measurements should be performed 
to determine the spectral function $S(E^\prime_x, \bd{p}_\alpha)$
that carries the information on the $\alpha$--momentum
distribution in the halo nucleus, or equivalently on the dineutron 
residual system. Furthermore, in the $\omega \rightarrow 0$ limit
elastic scattering from $^6$He will provide the most stringent test
of the charge distribution in $^6$He. If the present halo picture
is correct the charge distribution in $^6$He must be dictated by that
in the $\alpha$--core.

On the contrary, backward scattering is more favorable to measure
the halo--neutron peak. In particular, $\theta_e \gtrsim 160^\circ$,
50 $\lesssim E_0 \lesssim$ 100 MeV and $\omega < $ 40 MeV define
a region where the inclusive cross section is sizeable and dominated
by the halo neutron peak. 
Thus coincidence $(e,e^\prime n)$ measurements in this region will
allow to determine the spectral function $S(E^\prime_x, \bd{p}_n)$
that carries the information on the halo neutron momentum distribution, 
or equivalently on the unbound $^5$He residual system. 

The procedure used here will be extended to other halo nuclei in the
near future.

This work was supported by DGESIC (Spain) under contract
number PB98--0676.

%
%
%
%
\end{document}